# Using Query Mediators for Distributed Searching in Federated Digital Libraries

Naomi Dushay
Dept. of Computer Science
Cornell University
Ithaca, NY 14853-7501
naomi@cs.cornell.edu

James C. French
Dept. of Computer Science
University of Virginia
Charlottesville, VA 22903
french@cs.virginia.edu

Carl Lagoze
Dept. of Computer Science
Cornell University
Ithaca, NY 14853-7501
lagoze@cs.cornell.edu

**Abstract**

We describe an architecture and investigate the characteristics of distributed searching in federated digital libraries. We introduce the notion of a *query mediator* as a digital library service responsible for selecting among available search engines, routing queries to those search engines, and aggregating results. We examine operational data from the NCSTRL distributed digital library that reveals a number of characteristics of distributed resource discovery. These include availability and response time of indexers and the distinction between the query mediator view of these characteristics and the indexer view.

## 1 Introduction

Users of networked information systems expect rapid and accurate response to their queries. Fulfilling this requirement in the rapidly expanding World Wide Web has become increasingly problematic. Resource discovery tools, exemplified by the many well known "web crawlers", increasingly confront a number of problems including scalability, lack of domain specificity, and intellectual property restrictions [8]. These problems result, in part, from the centralized (albeit replicated) architecture of these tools.

Distributing query processing among a set of decentralized search engines offers a promising solution to these problems. Queries can be processed in parallel, reducing scalability issues. The functionality of individual search engines can be tailored to the needs of specific clientele and collections. Rights management issues can be handled through licensing agreements between specific search engines and specific sets of documents.

However, this distributed architecture requires sophisticated mechanisms to work properly and reliably. These include selectively distributing queries among the set of available search engines. This selection can be based both on the information content of the search engines and other factors including load, cost, licensing agreements, network latency, and server reliability.

We have been investigating these distributed resource discovery issues in the context of our broader research into federated digital library architecture. This architecture builds digital libraries from sets of individual services, each with defined functionality, which communicate among themselves using open protocols. The advantages of such an architecture include scalability and easy extension of functionality.

In this paper we pay particular attention to mechanisms for query distribution with regard to search engine performance. The question we attempt to answer is: *given a set of search engines that index the same information, which search engines can be* *selected to ensure rapid and reliable response?* As we will show, this consideration must take into account both the performance of the engines themselves and of the networks that are used to connect to them. In order to characterize the problem, we introduce the notion of a *query mediator*, which is a digital library service that is responsible for selectively distributing queries to search engines and accumulating their results. It accomplishes this task by collecting and processing metadata from search engines and other sources.

This paper is organized as follows. Section 2 describes the logical components of distributed searching architecture and defines the *query mediator* (QM) component. Section 3 describes characteristics of indexer performance and gives quantified findings of some of these characteristics, especially from the QM-view. In section 4 we examine the efficacy of different QM-view prediction methods for indexer performance. Section 5 outlines future research and section 6 has our conclusions.

## 2 An architecture for distributed searching

A federated digital library architecture [14] allows for semi-autonomous management of distributed services: participants retain autonomy over their services while using a common protocol to communicate with other services in the digital library. Two services in this model are *repositories*, which store and access digital documents, and *indexers*, which index information (metadata or full text) for digital documents and process queries on that information.

The information indexed in a particular indexer may be a replicate of that at other indexers, may be completely disjoint, or it may overlap in various ways with the indexed information at other indexers. The particular configuration depends on a mixture of administrative decisions, fault tolerance concerns and rights management issues. This creates a need for a mechanism to distribute queries to indexers with respect both to content (choosing indexers that have information relevant to the query) and to other factors such as cost or performance (choosing among multiple indexers indexing the same information).

This query distribution is one step of the multi-step resource discovery process described in Figure 1. First the user formulates a query. Next, the query is translated in preparation for indexer processing, after which it is distributed to a set of indexers. Results from these indexers are merged and finally formatted for presentation and delivered to the user.

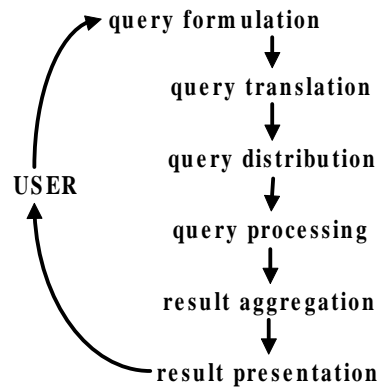

**Figure 1 – Resource discovery in a digital library**

The issue of query distribution has been examined both the database community and the digital library community. Historically, the database community has focused on the optimal distribution of indexing information across LANs and controlled WANs [2]. Areas that have received attention within the digital library community include content summarization for query distribution [7], protocols for meta-searching and metadata collection [1, 6].

Our work has focused on query distribution with respect to the performance of networks and indexers. In earlier work [10], we described how to use *collection services* and *connectivity regions* to address resource discovery in a globally distributed digital library. Connectivity regions are defined as a group of nodes on the network that among them have low latencies and a resistance to failure relative to nodes outside of the region. Metadata describing the apportioning of indexers to regions and the mapping of repositories to indexers is part of the collection service [9] and is used to determine indexer choices for query distribution: the query is only distributed to indexers within the connectivity region.

It is useful to think of query distribution as but one component of general class of functions called *query mediation*. In this sense, the notion of a separate *query mediator* service in a distributed digital library is a useful one. This builds on the notion of a *query router* introduced in [9]. Functionally, the QM stands between the user interface and indexer services. The user interface (UI) performs those digital library functions pertaining to direct user interaction, such as query formulations and result presentation, while the indexers do query processing. A query mediator (QM) does the remaining tasks (query translation, query distribution, result aggregation).

We note that while in conventional operation, the QM is effectively an intermediary between UIs and indexers, it is an independent service. Thus, its functionality is exposed for other types of agents and services.

In summary, QMs do the following:

- *Translate queries*. The QM receives queries from the UI and translates them into protocols understood by the indexers. Distributed digital libraries could have multiple protocols for communicating with indexers.

- *Choose indexers for query processing*. This choice may be based on any of the factors described above (e.g. content, performance …)

- *Adaptively react to operational conditions*. For example, if an indexer fails or doesn't respond by the search timeout, then one possibility is for the QM to send the query to an alternate indexer, providing fault tolerant distributed searching to the UI.

- *Aggregate search results*. The QM accumulates the results from all the indexers polled, merges them and removes any duplicates. It then sends the search results and any associated metadata about those results to the UI.

Choosing indexers and responding to failure requires the collection of information about indexer performance and other factors from various sources. The sources and characterizations of this metadata can be described as follows:

- *metadata from indexers*. This might include information about indexer content, indexer rights management, indexer pricing, or indexer load.

- *metadata accumulated and aggregated by the query mediators*. This includes search timeout values to inform the QM how long to wait for a response from a particular indexer. It also might include information about network connectivity and latency between the QM and a particular indexer.

- *metadata from the collection service*. The collection service provides information about what is included in the collection and how it is organized. So this metadata might include mappings of repositories to indexers, information about connectivity regions or other metadata relevant to the QM but administered at a collection wide level.

If the QM performs well, it will speedily deliver complete search results to the user. Similarly, a poor choice of search timeouts, poor indexer choices, or poor adaptations to operational conditions could result in slow or incomplete search results delivered to the digital library user.

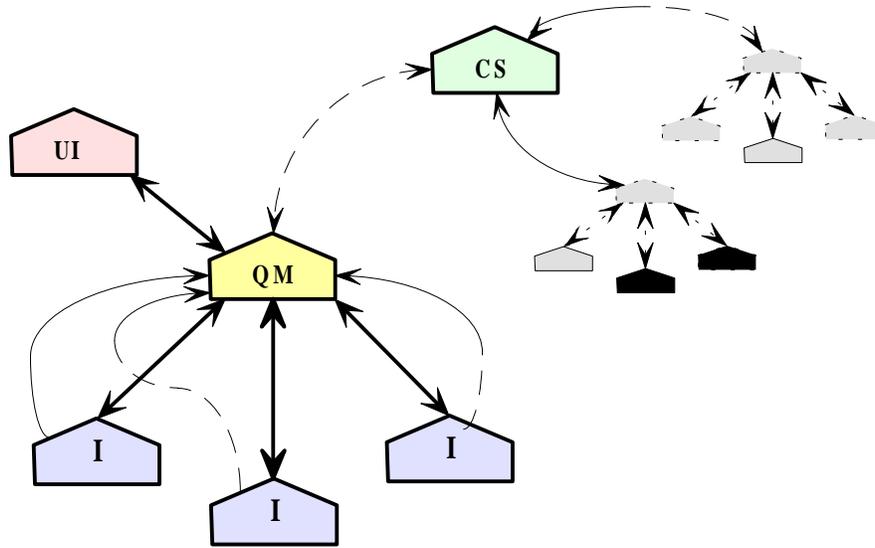

**Figure 2 - Interaction of services and metadata in distributed searching**

Figure 2 summarizes both the functional and metadata flows among the services described above. The services (depicted as chevrons) in the figure are labeled as follows: UI indicates a user interface, QM indicates a query mediator, I indicates indexers, and CS indicates a collection service. The solid lines show interactions between services during the processing of a query. As shown, the UI submits a query to the QM, which then distributes it to a set of indexers. These then return query results, which are merged by the QM, and returned to the UI. The curved dashed lines show the flow of metadata among the components. As shown, the QM accumulates metadata about the performance about each of the indexers. The QM exchanges metadata with the CS, which aggregates metadata from this QM and other QMs in the collection.

## 2.1 NCSTRL – a distributed digital library testbed

The distributed digital library on which we base our research is the Networked Computer Science Technical Reference Library (NCSTRL - pronounced "ancestral"). NCSTRL is an operational digital library employing a distributed, component-based architecture. The NCSTRL collection is globally distributed and made available through the Dienst [12] federated digital library architecture. Dienst is an open architecture and protocol [3] for distributed digital libraries that was developed as part of the DARPA-funded Computer Science Technical Reports Project[1]. These characteristics – global distribution, open interface, and production availability – make NCSTRL an ideal testbed for distributed digital library research (indeed, NCSTRL is one of the collections in the DARPA-funded Distributed Integration Testbed[2]).

The NCSTRL collection consists of institutions, or publishing organizations, each of which (at a minimum) provides a repository of digital documents and descriptive metadata [13] for those documents. These institutions are a combination of Ph.D. granting computer science departments, ePrint repositories, electronic journals, and research institutions.

---

[1] http://www.cnri.reston.va.us/cstr.html

[2] http://www.cnri.reston.va.us/integration-testbed.html

At the time of publication of this paper (February 1999), there were over 100 NCSTRL repositories and approximately 50 NCSTRL indexers worldwide.

The Dienst architecture specifies the operational characteristics of semi-autonomous core digital library services, as well as describing an open, extensible protocol for communicating among and with these digital library services. Core services include repositories, indexers and user interface gateways, as well as *collection services* which provide the mechanisms for federating these and other services into a digital library. While not formally defined in Dienst as a separate digital library service, the functionality of the QM is present in NCSTRL. Our research indicates that defining the QM as a separate service would be useful.

Each Dienst server, which implements and provides protocol access to a set of services, maintains logs containing operational and statistics messages. In order to study QM and indexer performance in an operational distributed digital library, we analyzed Dienst logs from the following five NCSTRL servers for the period from March 1, 1997 through April 30, 1997:

1. `NCSTRL` – the home page of NCSTRL, located at Cornell University.

2. `CS-TR` – the Cornell University Department of Computer Science Dienst server.

3. `LITE` – the Dienst server at the University of Virginia.

4. `BERKELEY` – the University of California at Berkeley Dienst server.

5. `FORTH` – the Institute of Computer Science, Foundation for Research and Technology – Hellas (ICS-FORTH) Dienst server.

Full details of how the logs were analyzed can be found in [4].

## 3   QM-view characterizations of indexer performance

We start this section by describing our approach to analyzing indexer performance data. Next we compare indexer performance from two perspectives and finally we examine indexer performance data from the perspective of the QM.

An indexer's performance does not appear the same to a QM as it appears to the indexer itself. To distinguish these two perspectives, we refer to the indexer view of itself as the *indexer-view* and define the *QM-view* as the QM's perspective on an indexer's performance.

In [4], we found that on average QMs spend 44-54% of their time waiting for indexers to respond. In order to reduce wait time for indexers, QMs must route queries to indexers that will respond quickly and reliably. However, determining which indexers fit that description is difficult. Indexer reliability and processing speed depend on factors such as indexer hardware characteristics, size of the indexes and current CPU load. But indexer performance from the QM-view depends on additional factors that are hard to predict, such as network connectivity and network loads. We would like the QM to have accurate predictions of indexer behavior to inform query routing decisions, but to make those predictions, we must first examine indexer behavior from both the indexer-view and the QM-view.

Our analysis focused on two key aspects of indexer performance:

- *Availability*: *whether or not the indexer responded within a time limit* (such as a search timeout). Whether or not an indexer responds to a QM is dependent on how long the QM listens for a response, in addition to network conditions, indexer CPU load, etc.

- *Response time*: how quickly the indexer responded, given that it responded. This is the elapsed time between the moment the query is sent from the QM and the moment results are received at the QM from the indexer.

Note that this characterization of indexer performance does not describe Internet topology or the network load; it merely focuses on the information a QM would use to choose among replicated or overlapped indexers.

**Table 1 - QM and indexer-views of mean response times for CS-TR indexer**

| calling QR | indexer view | QR view |
|---|---|---|
| cs-tr | 0.1 | 1.9 |
| ncstrl | 0.1 | 5.1 |
| berkeley | 0.1 | 5.1 |
| lite | 0.1 | 2.4 |

Table 1 shows a pronounced distinction between the indexer and QM-views of the mean response time of the CS-TR indexer. The indexer-view mean response time for all calling QMs is the same. However, the mean indexer response time from the viewpoint of each QM ranges from 1.9 seconds for the CS-TR QM to 5.1 seconds for

**Table 2 – QM-view of indexer response ratios between studied servers**

|  | cs-tr | | ncstrl* | | berkeley | | lite | | forth | |
|---|---|---|---|---|---|---|---|---|---|---|
|  | QR cs-tr sees indexer | indexer cs-tr as seen by QR | QR ncstrl* sees indexer | indexer ncstrl* as seen by QR | QR berkeley sees indexer | indexer berkeley as seen by QR | QR lite sees indexer | indexer lite as seen by QR | QR forth sees indexer | indexer forth as seen by QR |
| cs-tr | 0.99 | 0.99 | 0.99 | 0.87 | 0.95 | - | 0.99 | 0.53 | - | 0.89 |
| ncstrl* | 0.87 | 0.99 | 0.60 | 0.60 | 0.28 | - | 0.97 | 0.37 | - | - |
| berkeley | - | 0.95 | - | 0.28 | 0.07 | 0.07 | - | 0.32 | - | 0.79 |
| lite | 0.53 | 0.99 | 0.37 | 0.97 | 0.32 | - | 0.62 | 0.62 | - | 0.79 |
| forth | 0.89 | - | - | - | 0.79 | - | 0.79 | - | 0.97 | 0.97 |

### 3.1 QM-view vs. indexer-view of NCSTRL indexer performance

Our ability to examine the indexer-view of availability was constrained by Dienst logs that were not designed for this purpose; we therefore restrict our comparison to indexer response time from these two perspectives.

Table 1 compares the indexer-view of the mean query processing time of the CS-TR indexer to the QM-view of the mean indexer response time perceived by the four QMs in our study that contacted the CS-TR indexer: CS-TR, NCSTRL, BERKELEY and LITE.

the NCSTRL and BERKELEY QMs. The fastest of the QM-view mean response times is still 1800% greater than the indexer-view measurement, and it includes no network travel time as the CS-TR QM is located on the same machine as the CS-TR indexer. In this case, the discrepancy between QM-view and indexer-view measurements is solely due to indexer overhead not included in the indexer-view data and the time for the indexer to receive an *http* protocol request and to send a response back via *http* to the calling QM.

### 3.2 QM-view characterizations of NCSTRL indexer performance

To measure the QM-view of indexer availability, we studied the ratio between the number of attempts a QM made to contact an indexer and the number of responses the QM received from that indexer before the search timeout; this data is summarized in Table 2. To measure the QM-view of indexer response time, we studied indexer response times as culled from Dienst logs shows the indexer response ratios (Table 2) or mean indexer response times (Table 3) for the CS·TR indexer, as perceived by the QMs indicated. For example, in Table 3, the mean response time for the CS·TR indexer as perceived by the NCSTRL QM was 5.1

Table 3 - QM-view of mean indexer response times between studied servers

|  | QR cs-tr sees indexer | indexer cs-tr as seen by QR | QR ncstrl* sees indexer | indexer ncstrl* as seen by QR | QR berkeley sees indexer | indexer berkeley as seen by QR | QR lite sees indexer | indexer lite as seen by QR | QR forth sees indexer | indexer forth as seen by QR |
|---|---|---|---|---|---|---|---|---|---|---|
| cs-tr | 1.9 | 1.9 | 5.1 | 4.4 | 5.1 | - | 2.4 | 7.2 | - | 5.5 |
| ncstrl* | 4.4 | 5.1 | 7.9 | 7.9 | 7.9 | - | 3.8 | 9.8 | - | - |
| berkeley | - | 5.1 | - | 7.9 | 7.7 | 7.7 | - | 10.6 | - | 10.9 |
| lite | 7.2 | 2.4 | 9.8 | 3.8 | 10.6 | - | 10.2 | 10.2 | - | 6.6 |
| forth | 5.5 | - | - | - | 10.9 | - | 6.6 | - | 9.8 | 9.8 |

for the five QMs in our study; this data is summarized in Table 3.

Tables 2 and 3 show the QM-view of indexer response ratios (availability) and mean indexer response times, respectively, for indexers at the same Dienst servers as the five QMs in our study. Note that in Tables 2 and 3, "ncstrl" is marked with an asterisk because the NCSTRL QM and indexer are on different ports on the same host machine.

The data columns in Tables 2 and 3 are shown in five groupings of two columns each – one grouping for each site in our study. The meaning of the two columns within each grouping is as follows, using the first grouping as an illustrative example. The first data column, headed "QM cs-tr sees indexer," shows the indexer response ratios (Table 2) or mean indexer response times (Table 3) as perceived by the CS·TR QM for each of the indexers in our study. For example, in Table 2, the ratio of responses to failures for the NCSTRL indexer as perceived by the CS·TR QM was 0.87: out of 100 attempts to contact the NCSTRL indexer, 87 responded before the search timeout. The second data column, headed "indexer cs-tr as seen by QM," shows the seconds. The double column groupings for other sites in our study are derived similarly. Missing values in Tables 2 or 3 occur when a QM didn't call a particular indexer, or (similarly) when an indexer wasn't called by a particular QM. For example, the BERKELEY QM contacts all five indexers, but the BERKELEY indexer is only called by the BERKELEY QM.

There are three key observations that can be made from the data in Tables 2 and 3.

1. *The QM-view of indexer performance is unique for each indexer.* For example, in Table 2 the CS·TR QM has a response ratio of 0.99 for the CS·TR indexer, and a response ratio of 0.53 for the LITE indexer. Similarly, in Table 3 the CS·TR QM has a mean response time of 1.9 seconds for the CS·TR indexer, and a mean response time of 7.2 seconds for the LITE indexer.

2. *The same indexers appear different to different QMs.* For example, in Table 2 the NCSTRL indexer has a response ratio of 0.28 for the BERKELEY QM, and a response ratio of 0.97 for the LITE QM. Similarly, in Table 3 the FORTH indexer has a mean response time of 5.5 seconds for the CS-TR QM, and a mean response time of 10.9 seconds for the BERKELEY QM.

3. *Network topology is not sufficient to account for the QM-view of indexer behavior.*

   - *A QM doesn't always find the indexer on a local machine to be the best performer.* Table 2 shows that the BERKELEY and LITE QMs view their own indexers as the least reliable of any of the servers in our study. Table 3 shows that the NCSTRL, BERKELEY and LITE QMs all view the CS-TR indexer as fastest indexer in our study.

   - *The server interactions can be asymmetric with respect to performance.* That is, the response time perceived by QM A when accessing indexer B can be quite different from the response time observed by QM B when accessing indexer A. For example, Table 2 indicates the CS-TR QM recorded a response ratio of 0.53 for the LITE indexer, while the LITE QM recorded a ratio of 0.99 for the CS-TR indexer. The network connectivity between LITE and CS-TR must be good, otherwise the LITE QM could not have a ratio of 0.99 for the CS-TR indexer. But the low ratio for the CS-TR QM-view of the LITE indexer implies poor reliability or performance of the LITE indexer, given that the network connection is good. This is born out by the low indexer response ratios reported for the LITE indexer by all QMs.

   - *Indexer response times from the QM-view do not necessarily increase with network distance.* Table 3 indicates that the NCSTRL and BERKELEY QMs have the same mean indexer response times for the CS-TR and NCSTRL indexers. Given that the NCSTRL Dienst server is located at Cornell University in Ithaca, New York and the BERKELEY Dienst server is located in Berkeley, California, we would not expect these mean response times to be the same. It is even more surprising to note that the CS-TR Dienst server is also located at Cornell University in Ithaca, New York, but the NCSTRL QM response times are most similar to the distant BERKELEY QM. Another example of how response times are not closely related to Internet topology: in Table 3, the CS-TR, NCSTRL, BERKELEY and LITE QMs all see CS-TR as the fastest indexer, despite the network travel required by all but the CS-TR QM. So indexers located nearby on the Internet can be slower than indexers located further away.

These observations imply that sharing performance data among QMs will not be useful for accurately characterizing indexer behavior or for accurately predicting indexer performance. Instead, each QM should maintain its own idea of the operational environment: each QM must monitor individual indexer performance as part of its routine activity.

Since each QM has a unique view and must maintain its own unique metadata, and QMs as a group are responsible for particular resource discovery functionality as described in section 2, we assert the utility of the QM as a separate service in a distributed digital library.

## 4  QM-view predictions of indexer performance

As mentioned in previous sections, we would like to improve the QM mechanism for choosing among overlapped or replicated indexers. If we could accurately predict indexer performance from the QM-view, then we could optimize the choice of indexers made by the QM, reducing QM wait time and thus, user wait time.

QM-view indexer predictions need to address the following questions:

1. *Availability* -- will the indexer respond before the search timeout?

2. *Response time* -- how quickly will the indexer respond, given that it responds at all?

Answering these predictive questions accurately would be easy if the indexer performance data followed an observable pattern. Unfortunately, we were unable to discern clear overall patterns for indexer availability or response time in our data. Since we couldn't perceive patterns in the data, we applied a variety of predictive methods to QM-view indexer data. All methods we used combined past performance data for a given indexer at a particular QM into a succinct record and had algorithms requiring little QM processing time.

We believed that newer data would be a better predictor of indexer behavior than older data, so some of our predictive methods decayed old observations exponentially, either with a straight averaging window (a *low pass filter*) or dependent on the time elapsed between observations (a *timed low pass filter*). We were surprised to learn that simpler prediction methods, such as a running average with a window size of one (*single last observation*), or a running average with maximum window size (*running average*) usually performed as well as the more complex algorithms. This held true even though we optimized constants for the more complex algorithms on the same data we used to make the predictions.

We evaluated our predictions by comparing them to observations using mean square error (MSE). We learned, unsurprisingly, that prediction accuracy is related to data consistency; the utility of our work is that it quantifies the accuracy of different prediction methods, as well as the consistency of observed data from an operational, world-distributed digital library. For complete details on how we computed our predictions and how we evaluated them, see [5].

## 4.1 Availability predictions

Availability data was recorded as a binary measurement: either the indexer responded before the timeout (value 1) or it did not (value 0). When we applied our algorithms to this data, they produced a number between 0 and 1, which we then rounded in order to get a predictive value. In other words, an availability prediction of 0.3 became an availability prediction of 0 (we predicted the indexer would not respond before the search timeout); a prediction of 0.7 became 1 (we predicted the indexer would respond before the timeout).

**Table 4 - MSE for all availability predictions by QM**

| QR | no obs | single last obs | running average | low pass filter | timed low pass filter | | |
|---|---|---|---|---|---|---|---|
| | | | | | tlpf value | method A | method B |
| cs-tr | 160,392 | 0.10 | 0.10 | 0.10 | 0.09 | 0.08 | 0.09 |
| ncstrl | 113,511 | 0.09 | 0.09 | 0.09 | 0.09 | 0.07 | 0.08 |
| berkeley | 69,483 | 0.12 | 0.13 | 0.12 | 0.13 | 0.11 | 0.11 |
| lite | 6,363 | 0.07 | 0.09 | 0.07 | 0.07 | 0.07 | 0.07 |
| forth | 743 | 0.13 | 0.16 | 0.13 | 0.13 | 0.12 | 0.13 |

Since both the availability data and the availability predictions had binary values of 0 and 1, the error for any observation could be 0, 1, or –1. This means that the MSE is the same as the mean of the absolute value of the error: a MSE of .10 implies that one out of ten predictions was incorrect. We present the MSE for all availability predictions for each QM in our study in Table 4.

Table 4 shows the MSE for all availability predictions for any method ranges from 0.07 in a variety of prediction methods for the LITE QM to 0.16 for the running average prediction method for the FORTH QM. Given that the MSE for availability predictions could range from 0 to 1 mathematically, this is a small range, and the worst value implies that on average, only 16 out of 100 availability predictions were incorrect for the FORTH QM. Given that the bulk of the predictions were from the CS·TR and NCSTRL QMs, showing maximum MSE values of 0.10, we can say that approximately 90% of our indexer

## 4.2 Response time predictions

Table 5 shows the MSE for all indexer response time predictions for each QM in our study.

The MSE for all indexer response time predictions has a wide range: from 5.9 for the timed low pass filter - tlpf value prediction method for the CS·TR QM to 118.8 for the single last observation prediction method for the FORTH QM. Unlike the availability prediction MSEs, the response time prediction MSEs are not similar for all prediction methods. The lowest MSE for all but the CS·TR QM was from timed low pass filter prediction method B. However, all of these minimum MSE values were within 0.7 of the running average MSE values, except for the FORTH QM, which has far fewer observations. As with the availability predictions, a simple prediction method requiring no optimization

Table 5 - MSE for all response time predictions by QM

| QR | no obs | single last obs | running average | low pass filter | timed low pass filter | | |
|---|---|---|---|---|---|---|---|
| | | | | | tlpf value | method A | method B |
| cs-tr | 143,427 | 8.5 | 6.6 | 8.1 | 5.9 | 6.3 | 6.0 |
| ncstrl | 99,406 | 10.2 | 7.7 | 9.7 | 7.6 | 7.5 | 7.5 |
| berkeley | 54,852 | 24.7 | 15.2 | 23.4 | 15.1 | 15.2 | 15.0 |
| lite | 5,811 | 14.8 | 9.9 | 14.0 | 9.9 | 9.8 | 9.2 |
| forth | 479 | 118.8 | 69.4 | 113.2 | 67.1 | 69.2 | 64.0 |

availability predictions for any QM were accurate regardless of the predictive method used.

We surmise that any pattern, or lack thereof, in the availability data affects all predictive methods similarly, since MSE for any method for a given QM is at most 0.04 different from MSE for any other methods for the same QM. We do note that the timed low pass filter methods A and B (see [5] for a detailed explanation of these methods) have slightly lower MSE than any of the simpler methods, but the single last observation method performs nearly as well and requires no optimizing of constants.

of constants works nearly as well as more complex methods.

If we take the root mean square (RMS), or the square root of the MSE, then we get a rough approximation of the average error: on average, over all predictions for all indexers, running average predictions are roughly within SQRT (69.4) = 8.4 seconds of response time observations.

We now examine MSE for response time predictions for individual indexers at each QM, rather than MSE for all indexer predictions combined at each QM. Table 6 shows the highest MSE for an individual indexer at each QM in our study for each predictive method.

The lowest maximum MSE values for individual indexers are from the timed low pass filter prediction method B. As in Table 5, the simpler running average prediction method does nearly as well, although there is a difference of 8.6 between the running average MSE and timed low pass filter method B MSE for the maximum individual indexer MSE for the FORTH QM. We assert again that the simplicity of the running average prediction method probably outweighs the improved predictions from the timed low pass filter method B, given the complex computations required to optimize that method.

## 5 Future work

Having shown the utility of defining *query mediators* as a separate service, we intend to incorporate the QM service into the next generation digital library software currently under development [11]. We have been experimenting with STARTS [6] as a protocol for obtaining content information about both like and unlike indexers, and have been extending the STARTS protocol to accommodate performance metadata. In addition, we intend to improve the QM functionality NCSTRL by providing better monitoring of QM-view indexer performance. In both of these cases, we will use the expanded information about indexer content, performance, licensing and other

**Table 6 – Maximum MSE for QM response time predictions for individual indexers**

| QR | single last obs | running average | low pass filter | timed low pass filter | | |
|---|---|---|---|---|---|---|
| | | | | tlpf value | method A | method B |
| cs-tr | 65.7 | 51.9 | 63.1 | 51.3 | 51.9 | 45.2 |
| ncstrl | 30.3 | 21.1 | 28.6 | 21.0 | 21.1 | 21.0 |
| berkeley | 57.4 | 38.0 | 54.4 | 38.5 | 37.9 | 37.8 |
| lite | 50.2 | 23.7 | 47.7 | 23.0 | 23.7 | 23.0 |
| forth | 169.1 | 97.1 | 161.1 | 95.2 | 96.9 | 88.5 |

If we examine the RMS for predictions for individual indexers, then the running average method is within SQRT (97.1) = 9.9 seconds, on average, for the least accurate individual indexer predictions. In [5], we show that the accuracy of individual indexer predictions is related to the consistency of the response time observations. If we were to ignore the least consistent data, namely that from the FORTH QM, then the least accurate predictions using the running average method are within SQRT (51.9) = 7.2 seconds of the response time, on average. Another justification for ignoring the FORTH predictions is the small number of observations (see Table 5), though the number of observations does not correlate to the MSE value.

factors to influence indexer selection.

In addition, research has already been begun to address the need for more extensive and specific performance data in digital library server logs. This includes accumulated information on indexer reliability. One area of possible investigation is the definition of a standard log format that will serve as a tool for metrics and measurement comparisons among digital library systems.

## 6 Conclusions

In this paper, we defined the query mediator (QM) service and its role in distributed digital library architecture. QMs serve as intermediaries between the user interface and indexer services of digital libraries, translating queries to indexer protocols, choosing indexers to field the query, aggregating search results, and adapting to operational conditions. We explored the notion of how to measure indexer performance from the QM-view, and showed that each QM must monitor performance of individual indexers to have useful data for making performance predictions. This is because indexer behavior from the QM-view is not tied solely to indexer performance characteristics or to network topology.

We also discussed results of various QM-view algorithms for indexer availability and response time predictions and learned that complex algorithms that weighted recent indexer performance measurements more heavily than older measurements did not perform significantly better than simpler algorithms. We were able to predict indexer availability with 90% accuracy from a QM-view, on average for all indexers.

We also predicted QM-view response times for individual indexers within 7 seconds, or less, on average (depending on the consistency of the individual indexer data) using the running average method. Therefore, simple indexer performance monitoring by the QM should allow judicious choice of timeouts as well as reasonable predictions of whether or not an indexer will respond by the timeout. These improvements should improve QM functionality and hence improve the resource discovery process for users of distributed digital libraries.

## Acknowledgments

This work was supported by DARPA grant MDA 972-96-1-006 with the Corporation for National Research Initiatives, DARPA contract N66001-97-C-8542, and NSF grant CDA-9529253. This paper does not necessarily represent the views of CNRI, NSF, or DARPA. The authors are grateful to Elizabeth Slate and David Fielding especially for their contributions, and to UC Berkeley and ICS-FORTH for the log data.